\documentclass[graybox]{svmult}

\usepackage{type1cm}        
\usepackage{makeidx}         
\usepackage{graphicx}        
\usepackage{multicol}        
\usepackage[bottom]{footmisc}

\usepackage{newtxtext}       %
\usepackage{newtxmath}       

\makeindex             

\begin{document}

\title*{PHSD - a microscopic transport approach for strongly interacting systems}
\author{E.L.~Bratkovskaya, W.~Cassing, P. Moreau, L. Oliva, O.E. Soloveva, T. Song}
\institute{E.L.~Bratkovskaya \at GSI
Helmholtzzentrum f\"{u}r Schwerionenforschung GmbH, Darmstadt, Germany,
\at Institut f\"ur Theoretische Physik, %
 Goethe-Universit\"{a}t Frankfurt am Main, 
  Germany \\
  \email{elena.bratkovskaya@th.physik.uni-frankfurt.de}
\and W.~Cassing \at Institut f\"ur Theoretische Physik, %
 Justus-Liebig-Universit\"at Giessen,  Germany\\ \email{wolfgang.cassing@theo.physik.uni-giessen.de}
 \and P. Moreau, O.E. Soloveva \at Institut f\"ur Theoretische Physik, %
 Goethe-Universit\"{a}t Frankfurt am Main, 
  Germany
  \and L. Oliva \at Institut f\"ur Theoretische Physik, %
 Goethe-Universit\"{a}t Frankfurt am Main, 
  Germany
  \at GSI Helmholtzzentrum f\"{u}r Schwerionenforschung GmbH, Darmstadt, Germany
  \and  T. Song \at GSI Helmholtzzentrum f\"{u}r Schwerionenforschung GmbH, Darmstadt, Germany}
%
%
\maketitle

\abstract*{We present the basic ideas of the Parton-Hadron-String Dynamics
(PHSD) transport approach which is a microscopic covariant dynamical
model for strongly interacting systems formulated on the basis of
Kadanoff-Baym equations for Green's functions in phase-space representation (in 1st order gradient expansion beyond the quasi-particle approximation). The approach consistently describes the full evolution of a relativistic heavy-ion collision
from the initial hard scatterings and string formation through the dynamical deconfinement phase transition to the strongly-interacting quark-gluon plasma (sQGP)
as well as hadronization and the subsequent interactions in the expanding hadronic phase. The PHSD approach has been applied to p+p, p+A and A+A
collisions from lower SIS to LHC energies and been successful in describing a large number of experimental data including single-particle spectra, collective
flow and electromagnetic probes. Some highlights of recent PHSD results will be presented.}

\abstract{We present the basic ideas of the Parton-Hadron-String Dynamics
(PHSD) transport approach which is a microscopic covariant dynamical
model for strongly interacting systems formulated on the basis of
Kadanoff-Baym equations for Green's functions in phase-space representation (in 1st order gradient expansion beyond the quasi-particle approximation). The approach consistently describes the full evolution of a relativistic heavy-ion collision
from the initial hard scatterings and string formation through the dynamical deconfinement phase transition to the strongly-interacting quark-gluon plasma (sQGP)
as well as hadronization and the subsequent interactions in the expanding hadronic phase. The PHSD approach has been applied to p+p, p+A and A+A
collisions from lower SIS to LHC energies and been successful in describing a large number of experimental data including single-particle spectra, collective
flow and electromagnetic probes. Some highlights of recent PHSD results will be presented.}

\section{Introduction}
\label{sec:1}

The phase transition from partonic degrees of freedom (quarks and
gluons) to interacting hadrons is a central topic of modern high-energy
physics. In order to understand the dynamics and relevant scales of
this transition laboratory experiments under controlled conditions are
performed with relativistic nucleus-nucleus collisions.
Hadronic spectra and relative hadron abundances from these experiments
reflect  important aspects of the dynamics in the hot and dense zone
formed in the early phase of the reaction and collective flows provide information on the
transport properties of the medium generated on short time scales. Since relativistic heavy-ion collisions start with impinging nuclei in their groundstates a proper non-equilibrium description of the entire dynamics through possibly different phases up to the final asymptotic hadronic states - eventually showing some degree of equilibration - is mandatory.

About 40 years ago cascade calculations have been employed for the description of nucleus-nucleus collisions in the 1-2 AGeV range \cite{Cugnon} which provided already some good idea about the reaction dynamics including essentially nucleons, $\Delta$-resonances, pions and kaons. These calculations have been based on the Boltzmann equation which, however, is entirely classical and lacks quantum statistics appropriate for fermions and bosons. In particular the Pauli blocking for nucleons was found to be essential at lower bombarding energies and cascade calculations were extended in line with the Uehling-Ulenbeck equation for fermions \cite{UU} incorporating also some mean-field potential calculated in Hartree approximation with various two-body Skyrme forces. These type of transport models are denoted as Boltzmann-Uehling-Uhlenbeck (BUU) or Vlasov-Uehling-Uhlenbeck (VUU) models \cite{Bertsch,Cas90} and are still in use nowadays by some groups. Independently, Quantum Molecular Dynamical (QMD) models \cite{Aichelin} have been proposed in which the testparticles of the BUU/VUU approaches are replaced by Gaussians allowing for the simulation of single events while keeping the fluctuations. Explicit isospin degrees of freedom have been incorporated in IQMD \cite{Hartnack}, too. Since these type of models are based on a  Hamiltonian with fixed two-body forces one could evaluate the nuclear equation of state (EoS) at zero temperature or in thermal equilibrium and one of the primary issues was to extract the nuclear EoS from heavy-ion data by means of BUU/VUU or QMD calculations. Later on higher baryonic resonances as well as mesons like $\eta, K^\pm, K^0, {\bar K}^0, \rho, \omega, \phi$ have been incorporated which led to coupled-channel BUU (CBUU) approaches.

Apart from adding more hadronic degrees of freedom  in BUU/VUU fully relativistic formulations have been carried out on the basis of some Lagrangian density including a selected set of hadronic degrees of freedom \cite{Ko,Blaettel,Maruyama}. All baryons in such relativistic BUU (RBUU) models were propagated with scalar and vector selfenergies that were matched to reproduce collective flow data from heavy-ion collisions as well as particle spectra. This was a necessary step to go ahead in bombarding energy to ultra-relativistic p+A and A+A collisions, which were studied experimentally at the CERN SPS in the nineties. However, when increasing the number of degrees of freedom and adding high-mass short-lived resonances a lot of ambiguities entered the RBUU models since the couplings between the different hadronic species were unknown experimentally to a large extent. A way out was to incorporate the particle production by string formation and decay in line with the LUND model \cite{JETSET} which included only a formation time of hadrons ($\tau_F \approx 0.8$ fm/c) and a fragmentation function primarily fitted to hadron spectra from $e^+e^-$ annihilation, where only a single string is formed. Familiar versions are the Hadron-String-Dynamics (HSD) \cite{Ehehalt,HSD} or Ultra-relativistic Quantum Molecular Dynamics (UrQMD) \cite{UrQMD} approaches that have been applied to p+A and A+A reactions in a wide range of energies up to the top SPS energy of 158 AGeV. In fact, a direct comparison between these two models for p+p and A+A collisions has provided very similar results for hadron spectra and flows up to $\sqrt{s_{NN}}$ = 17.3 GeV \cite{Brat2004}. Furthermore, a relativistic extension of the QMD model - based on the NJL Lagrangian - has been proposed in Ref. \cite{Marty} but not followed up further except for a comparative study in Ref. \cite{Marty2}.

By Legendre transformations the Hamiltonian density could be easily evaluated in the RBUU models and the nuclear EoS in thermal (or chemical) equilibrium, accordingly. However, it was soon noticed that with increasing temperature $T$ and baryon density $\rho_B$ (or baryon chemical potential $\mu_B$) the energy density was likely to exceed some critical energy density ($\sim$ 1 GeV/fm$^3$) as indicated by early lattice QCD  (lQCD) calculations which also showed that with increasing $T$ a restoration of chiral symmetry should happen as seen from the temperature dependence of the scalar quark condensate $<{\bar q} q>(T)$. Furthermore, the interaction rates of strongly interacting hadrons reached a couple of hundred MeV at high baryon density $\rho_B$ and temperature $T$ such that the on-shell quasi-particle limit - applied in the standard models - became questionable. Furthermore, the spectral evolution especially of vector mesons in a hot and dense environment became of primary interest since the electromagnetic decay of vector mesons into dilepton pairs could be measured experimentally and was considered as a primary probe for the restoration of chiral symmetry in these media. To this end the relativistic transport approach was extended to off-shell dynamics on the basis of the Kadanoff-Baym dynamics in the turn of the Millenium \cite{Sascha1,Juchem,Knoll1} and it became possible to calculate the in-medium spectroscopy of vector mesons in heavy-ion collisions \cite{Brat2008}. On the other hand,  experimental observations at the
Relativistic Heavy Ion Collider (RHIC) indicated that a new
medium (Quark-Gluon Plasma (QGP)) was created in ultrarelativistic Au+Au collisions that is
interacting more strongly than hadronic matter.  Moreover, in line with theoretical studies in
Refs. \cite{Shuryak,Thoma,Andre} the QCD medium showed phenomena of an
almost perfect liquid of partons \cite{STARS,Miklos3} as extracted
from the strong radial expansion and the scaling of elliptic flow $v_2(p_T)$ of mesons and baryons with the number of constituent quarks and antiquarks \cite{STARS}.

The question about the  properties of this (nonperturbative) QGP liquid became of primary interest as well as
dynamical concepts describing the formation of color neutral hadrons from colored partons (hadronization). A fundamental
issue for hadronization  is the conservation of 4-momentum as well as the entropy problem because by fusion/coalescence of
massless (or low constituent mass) partons to color neutral bound states of low invariant mass (e.g. pions) the number of degrees of freedom and thus the total entropy is reduced in the hadronization process \cite{Koal1,Koal2,AMPT}. This problem - a violation of the second law of thermodynamics  as well as  the conservation of four-momentum and flavor currents - has been addressed in Ref. \cite{PRC08} on the basis of the Dynamical QuasiParticle Model (DQPM) employing covariant transition rates for the fusion of massive quarks and antiquarks to color neutral hadronic resonances or strings.
The DQPM is an effective field-theoretical model based on covariant propagators for quarks/antiquarks and gluons that have a finite width in their spectral functions (imaginary parts of the propagators). The determination/extraction of complex selfenergies for the partonic degrees of freedom has been performed in Refs.
\cite{Cassing06,Cassing07} by fitting lattice QCD (lQCD)
data within the DQPM and thus extracting a temperature-dependent effective coupling (squared) $g^2(T/T_c)$, where $T_c$ denotes the critical temperature for the phase transition from hadrons to partons. This transition at low baryon chemical potential was found to be a crossover and the critical temperature $T_c$ could be extrated from the lQCD data. In fact,
the DQPM allows for a simple and transparent interpretation of
lattice QCD results for thermodynamic quantities as well as
correlators and leads to effective strongly interacting partonic
quasiparticles with broad spectral functions.  For a review on
off-shell transport theory and results from the DQPM in comparison
to lQCD we refer the reader to Refs. \cite{Crev,Review}.

Now a consistent dynamical approach - valid also for strongly
interacting systems - could be formulated on the basis of
Kadanoff-Baym (KB) equations \cite{Sascha1} or off-shell
transport equations in phase-space representation, respectively
\cite{Sascha1,Juchem,Knoll1}. In the KB theory the field quanta
are described in terms of dressed propagators with complex selfenergies (as in the DQPM).
Whereas the real part of the selfenergies can be related to
mean-field potentials (of Lorentz scalar, vector or tensor type),
the imaginary parts  provide information about the lifetime and/or
reaction rates of time-like particles \cite{Crev}. Once the
proper (complex) selfenergies of the degrees of freedom are known
the time evolution of the system is fully governed  by off-shell
transport equations (as described in Refs. \cite{Sascha1,Crev}).

\section{The PHSD approach}
\label{sec:2}

The Parton-Hadron-String-Dynamics  approach is a microscopic
covariant transport model that incorporates effective partonic as
well as hadronic degrees of freedom and involves a dynamical
description of the hadronization process from partonic to hadronic
matter. Whereas the hadronic part is essentially equivalent to the
conventional HSD approach \cite{HSD} the partonic
dynamics is based on the Dynamical Quasiparticle Model
\cite{Cassing06,Cassing07} which
describes QCD properties in terms of single-particle Green's
functions in the form
\begin{equation} \label{prop}
G^R(\omega, {\bf p}) = \left(
\omega^2 - {\bf p}^2 - M^2 + 2i\gamma \omega \right)^{-1} , \end{equation} where $M$ denotes the (resummed) mass of the parton and $\gamma$ its width, while $(\omega, {\bf p})$ is the parton four-momentum.  With the (essentially three) DQPM parameters for the temperature-dependent effective coupling $g^2(T/T_c)$ fixed by
lattice QCD results the approach is fully defined in the partonic phase.
We mention in passing that the off-shell transport equations can be solved within an
extended testparticle Ansatz \cite{Sascha1,Crev}.

One might ask whether the quasiparticle properties -- fixed in
thermal equilibrium -- should be appropriate also for the
nonequilibrium configurations. This question is nontrivial and can
only be answered by detailed investigations e.g. on the basis of
Kadanoff-Baym equations. We recall that such studies have been
summarized in Ref.~\cite{Crev} for strongly interacting
scalar fields that initially are far off-equilibrium and simulate
momentum distributions of colliding systems at high relative
momentum. The results for the effective parameters $M$ and $\gamma$,
which correspond to the time-dependent pole mass and width of the
propagator (\ref{prop}), indicate that the quasiparticle properties - except for
the very early off-equilibrium configuration - are close to the
equilibrium mass and width even though the phase-space distribution
of the particles is far from equilibrium (cf. Figs. 8 to 10 in Ref.
\cite{Crev}). Accordingly, we will adopt the equilibrium
quasiparticle properties also for phase-space configurations out of
equilibrium as appearing in relativistic heavy-ion collisions. The
reader has to keep in mind that this approximation is well
motivated, however, not fully equivalent to the exact solution.

On the hadronic side PHSD includes explicitly the  baryon and antibaryon octet and
decouplet, the $0^-$- and $1^-$-meson nonets as well as selected
higher resonances as in HSD~\cite{HSD}. Hadrons of higher
masses ($>$ 1.5 GeV in case of baryons and $>$ 1.3 GeV in case of
mesons) are treated as "strings" (color-dipoles) that  decay to the
known (low-mass) hadrons according to the JETSET algorithm
\cite{JETSET}. We discard an explicit recapitulation of the string
formation and decay and refer the reader to the original work
\cite{JETSET}.

\subsection{Hadronization}
\label{sect:hadroniz}
Whereas the dynamics of partonic as well as hadronic systems is fixed by the DQPM or HSD, respectively,
the change in the degrees of freedom has to be specified in line with the lattice QCD equation of state.
The hadronization, i.e. the transition from partonic to hadronic
degrees of freedom, has been introduced in Refs. \cite{PRC08,CaBra09} and is repeated here for
completeness. The hadronization is implemented in PHSD by local
covariant transition rates e.g. for $q+\bar{q}$ fusion to a mesonic state $m$ of four-momentum
$p= (\omega, {\bf p})$ at space-time point $x=(t,{\bf x})$:
\begin{eqnarray}
&&\phantom{a}\hspace*{-5mm} \frac{d N_m(x,p)}{d^4x d^4p}= Tr_q
Tr_{\bar q} \
  \delta^4(p-p_q-p_{\bar q}) \
  \delta^4\left(\frac{x_q+x_{\bar q}}{2}-x\right) 
  \omega_q \ \rho_{q}(p_q)
   \  \omega_{\bar q} \ \rho_{{\bar q}}(p_{\bar q})
\nonumber \\
&& \times |v_{q\bar{q}}|^2 \ W_m(x_q-x_{\bar q},(p_q-p_{\bar q})/2)
\, \, N_q(x_q, p_q) \
  N_{\bar q}(x_{\bar q},p_{\bar q}) \ \delta({\rm flavor},\, {\rm color}).
\label{trans}
\end{eqnarray}
In Eq. (\ref{trans}) we have introduced the shorthand notation,
\begin{equation}
Tr_j = \sum_j \int d^4x_j \int \frac{d^4p_j}{(2\pi)^4} \ ,
\end{equation}
where $\sum_j$ denotes a summation over discrete quantum numbers
(spin, flavor, color); $N_j(x,p)$ is the phase-space density of
parton $j$ at space-time position $x$ and four-momentum $p$.  In Eq.
(\ref{trans}) $\delta({\rm flavor},\, {\rm color})$ stands
symbolically for the conservation of flavor quantum numbers as well
as color neutrality of the formed hadronic state $m$ which can be
viewed as a color-dipole or "pre-hadron".  Furthermore, $v_{q{\bar
q}}(\rho_p)$ is the effective quark-antiquark interaction  from the
DQPM  (displayed in Fig. 10 of Ref. \cite{Cassing07}) as a
function of the local parton ($q + \bar{q} +g$) density $\rho_p$ (or
energy density). Furthermore, $W_m(x,p)$ is the dimensionless
phase-space distribution of the formed "pre-hadron", i.e.
\begin{equation} \label{Dover:1991zn} W_m(\xi,p_\xi) =
\exp\left( \frac{\xi^2}{2 b^2} \right)\ \exp\left( 2 b^2 (p_\xi^2-
(M_q-M_{\bar q})^2/4) \right)
\end{equation} with $\xi = x_1-x_2 = x_q - x_{\bar q}$ and $p_\xi = (p_1-p_2)/2
= (p_q - p_{\bar q})/2$. The width parameter $b$ is
fixed by $\sqrt{\langle r^2 \rangle} = b$ = 0.66 fm (in the rest
frame) which corresponds to an average rms radius of mesons. We note
that the expression (\ref{Dover:1991zn}) corresponds to the limit of
independent harmonic oscillator states and that the final
hadron-formation rates are approximately independent of the
parameter $b$ within reasonable variations. By construction the
quantity (\ref{Dover:1991zn}) is Lorentz invariant; in the limit of
instantaneous hadron formation, i.e. $\xi^0=0$, it provides a
Gaussian dropping in the relative distance squared $({\bf r}_1 -
{\bf r}_2)^2$. The four-momentum dependence reads explicitly (except
for a factor $1/2$)
\begin{equation} (E_1 - E_2)^2 - ({\bf p}_1 - {\bf p}_2)^2 -
(M_1-M_2)^2 \leq 0
\end{equation} and leads to a negative argument of the second
exponential in Eq. (\ref{Dover:1991zn}) favoring the fusion of
partons with low relative momenta $p_q - p_{\bar q}= p_1-p_2$.

Some comments on the hadronization scheme are in order: The
probability for a quark to hadronize is essentially proportional to
the timestep $dt$ in the calculation, the number of possible
hadronization partners in the volume $dV \sim$ 5 fm$^3$ and the
transition matrix element squared (apart from the gaussian overlap
function). For temperatures above $T_c$ the probability is rather
small ($\ll$ 1) but for temperatures close to $T_c$ and below $T_c$
the matrix element  becomes very large since it essentially scales
with the effective coupling squared $g^2(T/T_c)$ which is strongly enhanced
in the infrared. For a finite timestep $dt$ -- as used in the
calculations -- the probability becomes larger than 1 which implies
that the quark has to hadronize with some of the potential
antiquarks in the actual timestep if the temperature or energy
density becomes too low. Furthermore, the gluons practically freeze
out close to $T_c$ since the mass difference between quarks and
gluons increases drastically with decreasing temperature and the
reaction channel $g \leftrightarrow q + {\bar q}$ is close to
equilibrium. This implies that all partons hadronize. Due to
numerics some leftover partons may occur at the end of the
calculations which are forced to hadronize by increasing the
volume $dV$ until they have found a suitable partner. In practice
the forced hadronization was only used for LHC energies where the
computational time was stopped at $\sim$ 1000 fm/c when partons with
rapidities close to projectile or target rapidity did not yet
hadronize due to time dilatation ($\gamma_{cm} \approx $ 1400).

Related transition rates  (\ref{trans}) are
defined for the fusion of three off-shell quarks ($q_1+q_2+q_3
\leftrightarrow B$) to a color neutral baryonic ($B$ or $\bar{B}$)
resonances of finite width (or strings) fulfilling energy and
momentum conservation as well as flavor current conservation (cf.
 Ref. \cite{CaBra09}). In contrast to the familiar
coalescence models  this hadronization scheme solves the problem of
simultaneously fulfilling all conservation laws and the constraint
of entropy production. For further details we refer the reader to
Refs. \cite{PRC08,CaBra09}.

\subsection{Initial conditions}
\label{sect:init}

The initial conditions for the parton/hadron dynamical system have
to be specified additionally.  In order to describe relativistic
heavy-ion reactions we start with two nuclei in their
semi-classical groundstate, boosted towards each other with a
velocity $\beta$ (in $z$-direction), fixed by the bombarding energy.
The initial phase-space distributions of the projectile and target
nuclei are determined in the local Thomas-Fermi limit as in the HSD
transport approach~\cite{HSD} or the UrQMD model
\cite{UrQMD}. We recall that at relativistic
energies the initial interactions of two nucleons are well described
by the excitation of two color-neutral strings which decay in time
to the known hadrons (mesons, baryons, antibaryons) \cite{JETSET}.
Initial hard processes - i.e. the short-range high-momentum transfer
reactions that can be well described by perturbative QCD - are
treated in PHSD (as in HSD) via PYTHIA. The novel
element in PHSD (relative to HSD) is the string melting concept as
also used in the AMPT model \cite{AMPT} in a similar context.
However, in PHSD the strings (or possibly formed hadrons) are only
allowed to melt if the local energy density $\epsilon(x)$ (in the
local rest frame) is above  the transition energy density
$\epsilon_c$ which in the  DQPM  is $\epsilon_c \approx 0.5
$ GeV/fm$^3$. The mesonic strings then decay to quark-antiquark
pairs according to an intrinsic quark momentum distribution,
\begin{equation} \label{mom0}
F({\bf q}) \sim \exp(- 2 b^2 {\bf q}^2) \ ,
\end{equation}
in the meson rest-frame (cf. Eq. (\ref{trans}) for the inverse
process). The parton final four-momenta are selected randomly
according to the momentum distribution (\ref{mom0}) (with $b$= 0.66
fm), and the parton-energy distribution is fixed by the DQPM at
given energy density $\epsilon(\rho_s)$ in the local cell with
scalar parton density $\rho_s$. The flavor content of the $q\bar{q}$
pair is fully determined by the flavor content of the initial
string. By construction the "string melting" to massive partons
conserves energy and momentum as well as the flavor content. In
contrast to Ref. \cite{AMPT} the partons are of finite mass --
in line with their local spectral function -- and obtain a random
color $c= (1,2,3)$ or $(r,b,g)$ in addition. Of course, the color
appointment is color neutral, i.e. when selecting a color $c$ for
the quark randomly the color for the antiquark is fixed by $-c$. The
baryonic strings melt analogously into a quark and a diquark while
the diquark, furthermore, decays to two quarks. Dressed gluons are
generated by the fusion of nearest neighbor $q+ {\bar q}$ pairs ($q+
{\bar q} \rightarrow g$) that are flavor neutral until the ratio of
gluons to quarks reaches the value $N_g/(N_q + N_{\bar q})$ given by
the DQPM for the energy density of the local cell. This
recombination is performed for all cells in space during the
passage time of the target and projectile (before the calculation
continues with the next timestep) and conserves the four-momentum as
well as the flavor currents. We note, however, that the initial
phase in PHSD is dominated by quark and antiquark
degrees of freedom\cite{Moreau}.

 Apart from proton-proton, proton-nucleus or
nucleus-nucleus collisions the PHSD approach can also be employed to
study the properties of the interacting hadron/parton system in a
finite box with periodic boundary conditions \cite{93}. To this aim the system
is initialized by a homogeneous distribution of test-particles in a
finite box with a momentum distribution close to a thermal one. Note
that in PHSD the system cannot directly be initialized by a
temperature and chemical potential since these Lagrange parameters
can only be determined when the system has reached a thermal and
chemical equilibrium, i.e. when all forward and backward reaction
rates have become equal; this is easy to check in the transport simulations.

\subsection{Partonic cross sections}
\begin{figure*}[h!]
	\centering
	\includegraphics[width=0.75\linewidth]{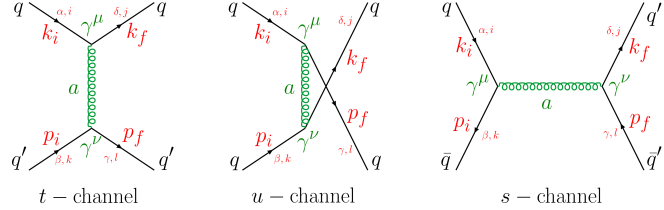}
	\caption{Leading order Feynman diagrams for the $q q' \rightarrow q q' $ and $q \bar{q} \rightarrow q' \bar{q}' $ processes. The initial and final 4-momenta are $k_i$ and $p_i$, and $k_f$ and $p_f$, respectively. The indices ${ i,j,k,l}=1-3$ denote the quark colors, ${ a}=1-8$ the gluon colors while the quark flavor is indicated by the indices ${
\alpha,\beta,\delta,\gamma}=u,d,s,...$.}
	\label{fig-feynman1}
\end{figure*}

On the partonic side the following elastic and inelastic
interactions are included in PHSD $qq \leftrightarrow qq$, $\bar{q}
\bar{q} \leftrightarrow \bar{q}\bar{q}$, $gg \leftrightarrow gg$,
$gg \leftrightarrow g$, $q\bar{q} \leftrightarrow g$, $q g \leftrightarrow q g$,
$g \bar{q} \leftrightarrow g \bar{q}$  exploiting
detailed-balance with cross sections calculated from the leading Feynman diagrams employing the effective propagators and couplings $g^2(T/T_c)$ from the DQPM \cite{Pierre}. As an example we show in Fig. \ref{fig-feynman1} the leading order Feynman diagrams for the $q q' \rightarrow q q' $ and $q \bar{q} \rightarrow q' \bar{q}' $ processes.

Partonic reactions such as $g+q \leftrightarrow q$
or $g+g \leftrightarrow q+ {\bar q}$  have been discarded
in the present calculations due to their low rates since
the large mass of the gluon leads to a strong
mismatch in the energy thresholds between the initial and final channels. In this case $q$ stands for the 4 lightest quarks ($u,d,s,c)$. Furthermore,    the evaluation of
photon and dilepton production is calculated perturbatively and channels
like $g+q \rightarrow  q+\gamma$ are included. In this case the probability for
photon (dilepton) production from each channel is added up and integrated over space and time \cite{Review}
without introducing any new parameter in the PHSD approach since the electromagnetic coupling is well known.

Numerical tests of the parton
dynamics with respect to conservation laws, interaction rates in and
out-off equilibrium in a finite box with periodic boundary conditions
have been presented in Ref. \cite{93}. In fact, in
Ref. \cite{93} it was shown that the PHSD
calculations in the box give practically the same results in
equilibrium as the DQPM. We note in passing that the total energy
is conserved in the box calculations up to about 3 digits while in
the heavy-ion collisions addressed here in the following the violation
of energy conservation is typically less than 1 \% \cite{CaBra09}.

\section{Transport properties of the partonic system}

The starting point to evaluate viscosity coefficients of partonic matter is the Kubo formalism \cite{85} which was also used to calculate the viscosities  within the PHSD in a box with periodic boundary conditions (cf. Ref. \cite{Ozvenchuk13}). We focus here on the calculation of the shear viscosity based on Ref. \cite{87} which reads:
\begin{equation}
\eta^{\text{Kubo}}(T,\mu_q)  = - \int \frac{d^4p}{(2\pi)^4}\ p_x^2 p_y^2 \sum_{i=q,\bar{q},g} d_i\ \frac{\partial f_i(\omega)}{\partial \omega}\ \rho_i(\omega,\mathbf{p})^2 \label{eta_Kubo} \end{equation} $$
=  \frac{1}{15T} \int  \frac{d^
4p}{(2\pi)^4}\ \mathbf{p}^4 \sum_{i=q,\bar{q},g} d_i \left( (1 \pm f_i(\omega)) f_i(\omega) \right) \rho_i(\omega,\mathbf{p})^2 ,
$$
where the notation $f_i(\omega) = f_i(\omega,T,\mu_q)$ is used for the distribution functions, and $\rho_i$ denotes the spectral function of the partons, while $d_i$ stand for the degeneracy factors. We note that the derivative of the distribution function accounts for the Pauli-blocking (-) and Bose-enhancement (+) factors. Following Ref. \cite{89}, we can evaluate the integral over $\omega = p_0$ in Eq. (\ref{eta_Kubo}) by using the residue theorem. When keeping only the leading order contribution in the width $\gamma(T,\mu_B)$ from the residue - evaluated at the poles of the spectral function $\omega_i = \pm \tilde{E}(\mathbf{p}) \pm i \gamma$ - we finally obtain:
\begin{equation} \label{eta_on}
 \eta^{\text{RTA}}(T,\mu_q)  = \frac{1}{15T} \int \frac{d^3p}{(2\pi)^3} \sum_{i=q,\bar{q},g}
   \frac{\mathbf{p}^4}{E_i^2 \ \Gamma_i(\mathbf{p},T,\mu_q)}\ d_i \left( (1 \pm f_i(E_i)) f_i(E_i) \right)   ,  \end{equation}
which corresponds to the expression derived in the relaxation-time approximation (RTA) \cite{94} by identifying the interaction rate $\Gamma$ with $2\gamma$ as expected from transport theory in the quasiparticle limit \cite{21}. We recall that $\gamma$ is the width parameter in the parton propagator (1). The interaction rate $\Gamma_i(\mathbf{p},T,\mu_q)$ (inverse relaxation time) is calculated microscopically from the collision integral using the differential cross sections for parton scattering as described in Section 2.3. We, furthermore,  recall that the pole energy is $E_i^2 = p^2 + M_i^2$ where $M_i$ is the pole mass given in the DQPM.
We use here the notation $\sum_{j=q,\bar{q},g}$ which includes the contribution from all possible partons which in our case are the gluons and the (anti-)quarks of three different flavors ($u,d,s$).

\begin{figure}
	\centering
	\includegraphics[width=0.5\linewidth]{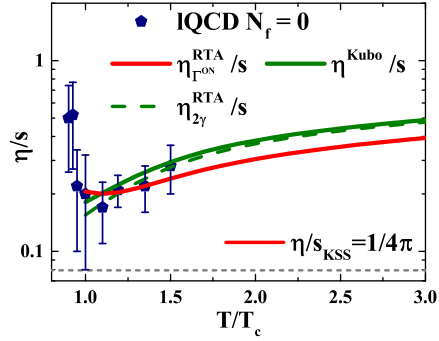}
    \caption{The ratio of shear viscosity to entropy density as a function of the scaled temperature $T/T_c$ for $\mu_B = 0$
        from Eq. (\ref{eta_Kubo}-\ref{eta_on}). The solid green line $(\eta^{\text{Kubo}}/s)$ shows the results from the original DQPM in the Kubo formalism while the dashed green line $(\eta^{\text{RTA}}_{2\gamma}/s)$ shows the same result in the quasiparticle approximation (\ref{eta_on}). The solid red line $(\eta^{\text{RTA}}_{\Gamma^{\text{on}}}/s)$ results from Eq. (\ref{eta_on}) using the interaction rate $\Gamma^{\text{on}}$  calculated by the microscopic differential cross sections in the on-shell limit. The dashed gray line demonstrates the Kovtun-Son-Starinets bound \cite{100} $(\eta/s)_{\text{KSS}} = 1/(4\pi)$, and the symbols show lQCD data for pure SU(3) gauge theory obtained within the Backus-Gilbert method taken from Ref. \cite{101} (pentagons).}
	\label{fig_eta}
\end{figure}

The actual results are displayed in Fig. \ref{fig_eta} for the
ratio of shear viscosity to entropy density $\eta/s$ as a function
of the scaled temperature $T/T_c$ for $\mu_B$ = 0 in comparison to
those from lattice QCD \cite{101}.
The solid green line $(\eta^{\text{Kubo}}/s)$ shows the result from the original DQPM in the
Kubo formalism  while the dashed green line $(\eta^{\text{RTA}}_{2\gamma}/s)$ shows the same result in
the relaxation-time approximation (\ref{eta_on}) by replacing
$\Gamma_i$ by $2\gamma_i$. The solid red line $(\eta^{\text{RTA}}_{\Gamma^{\text{on}}}/s)$ results from Eq.
(\ref{eta_on}) using the interaction rate $\Gamma^{\text{on}}$
 calculated by the microscopic differential cross
sections in the on-shell limit. We find that - apart from
temperatures close to $T_c$ -  the ratios $\eta/s$ do not differ
very much and have a similar behavior as a function of temperature.
The approximation (\ref{eta_on}) of the shear viscosity is found to
be very close to the one from the Kubo formalism (\ref{eta_Kubo})
indicating that the quasiparticle limit ($\gamma \ll M$) holds in
the DQPM.

	\begin{figure*}
		\centering
		\includegraphics[width=0.49\linewidth]{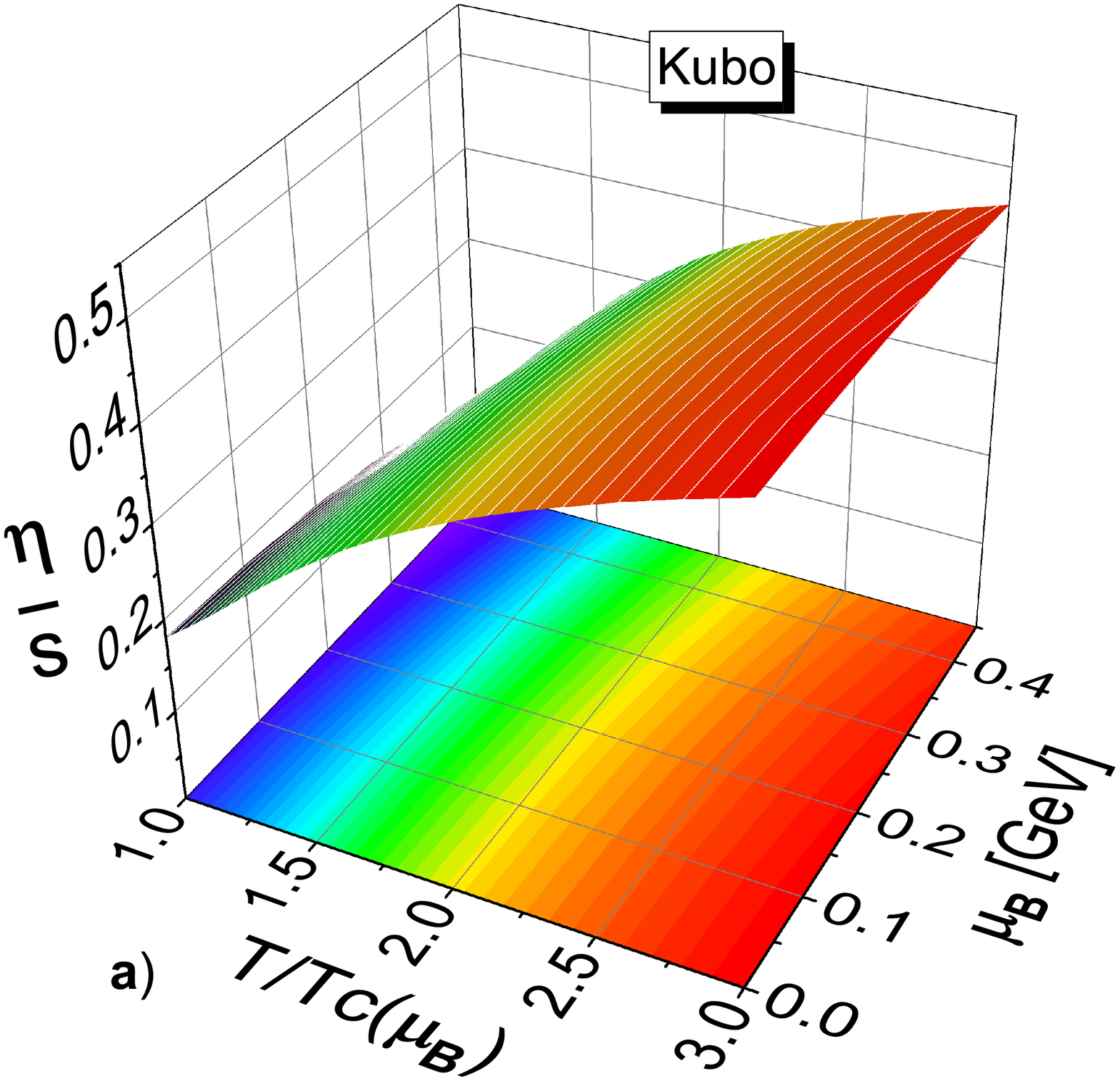}
		\includegraphics[width=0.49\linewidth]{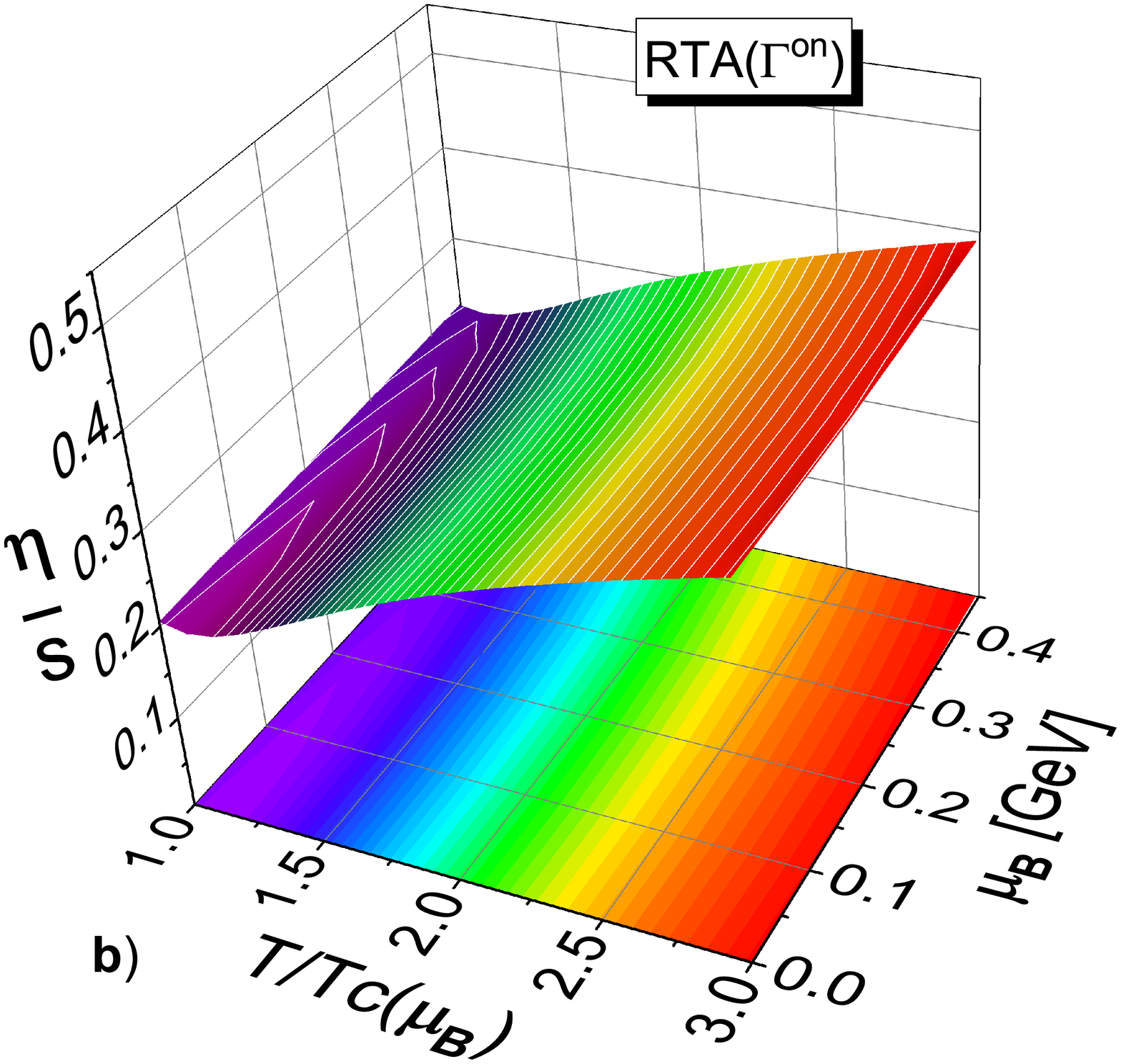}
		\caption{The ratio of shear viscosity to entropy density $\eta/s$ as a function of the scaled
			temperature $T/T_c(\mu_B)$ and baryon chemical potential $\mu_B$ calculated within the Kubo formalism (a) from Eq. (\ref{eta_Kubo}) and in the Relaxation Time Approximation (RTA) (b) from Eq. (\ref{eta_on}) using the on-shell interaction rate $\Gamma^{\text{on}}$.}
		\label{fig_eta_3d}
	\end{figure*}

An overview for the ratio of shear viscosity to entropy density $\eta/s$  as a function of the scaled temperature $T/T_c(\mu_B)$ and $\mu_B$ is given Fig. \ref{fig_eta_3d} in case of the Kubo formalism (a) (\ref{eta_Kubo}) and the relaxation-time approximation (\ref{eta_on}) (b). There is no strong variation with $\mu_B$ for fixed $T/T_c(\mu_B)$, however, the ratio increases slightly with $\mu_B$ in the on-shell limit while it slightly drops with $\mu_B$ in the Kubo formalism for the DQPM. Accordingly, there is some model uncertainty when extracting the shear viscosity in the different approximations.

In summarizing this section we find that the results for the ratio of shear viscosity over entropy density from the original DQPM and those from the microscopic calculations are  similar and within error bars compatible with present results from lattice QCD. However, having the differential cross sections for each partonic channel at hand one might find substantial differences for non-equilibrium configurations as encountered in relativistic heavy-ion collisions where a QGP is formed initially out-off equilibrium.

\section{Observables from relativistic nucleus-nucleus collisions}
\label{Section6}
We briefly report on results from PHSD calculations at lower and intermediated energies covered experimentally by the AGS (BNL) and SPS (CERN) with a focus on central Au+Au or Pb+Pb collisions. In this energy range the average baryon chemical potential $\mu_B$ is essentially finite - contrary to RHIC and LHC energies - and one might find some traces of the explicit $\mu_B$ dependence of the partonic cross sections in observables. To this end we compare results for the rapidity distributions from the PHSD calculations based on the default DQPM parameters (PHSD4.0) \cite{Palmese} with the new PHSD5.0 including the differential partonic cross sections for the individual partonic channels at finite $T$ and $\mu_B$ (cf. Ref. \cite{Pierre}). A comparison to the available experimental data is included (for orientation) but not discussed explicitly since this has been done in more detail in Ref. \cite{Palmese}. When implementing the differential cross sections and parton masses into the PHSD5.0 approach one has to specify the Lagrange parameters $T$ and $\mu_B$ in each computational cell in space-time. This has been done by employing the DQPM equation of state, which is practically identical to the lattice QCD equation of state,  and a diagonalization of the energy-momentum tensor from PHSD as described in Ref. \cite{Pierre}.

\begin{figure*}[thb]
	\centering
	\includegraphics[width=110mm]{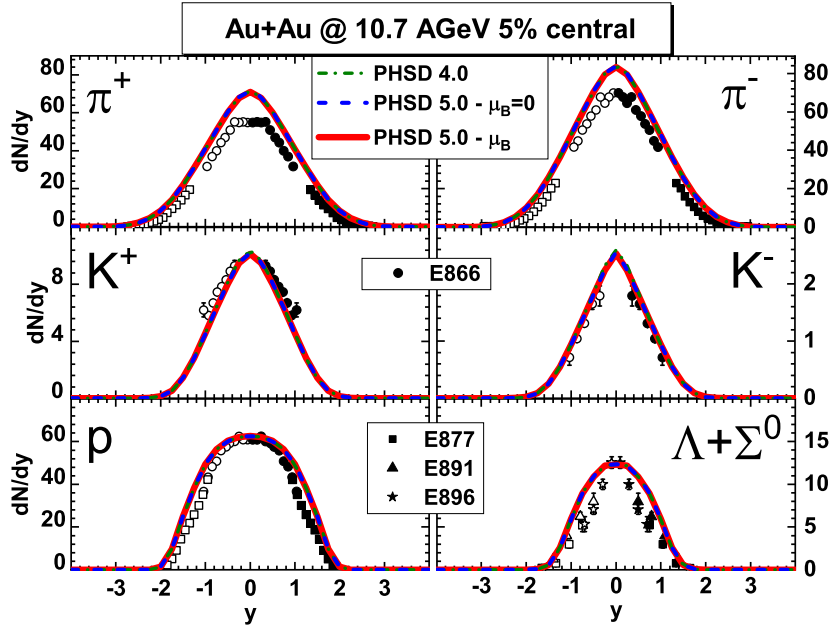}
	\caption{The rapidity distributions for  5\% central Au+Au collisions at 10.7 AGeV for PHSD4.0 (green dot-dashed lines), PHSD5.0 with partonic cross sections and parton masses calculated for $\mu_B$ = 0 (blue dashed lines) and with cross sections and parton masses evaluated at the actual chemical potential $\mu_B$ in each individual space-time cell (red lines) in comparison to the experimental data from the E866 \cite{110}, E877 \cite{111}, E891 \cite{112}, E877 \cite{113} and E896 \cite{114} collaborations. All PHSD results are the same within the linewidth.}
	\label{fig-dNdy-11AGeV}
\end{figure*}

\begin{figure*}[thb]
	\centering
	\includegraphics[width=110mm]{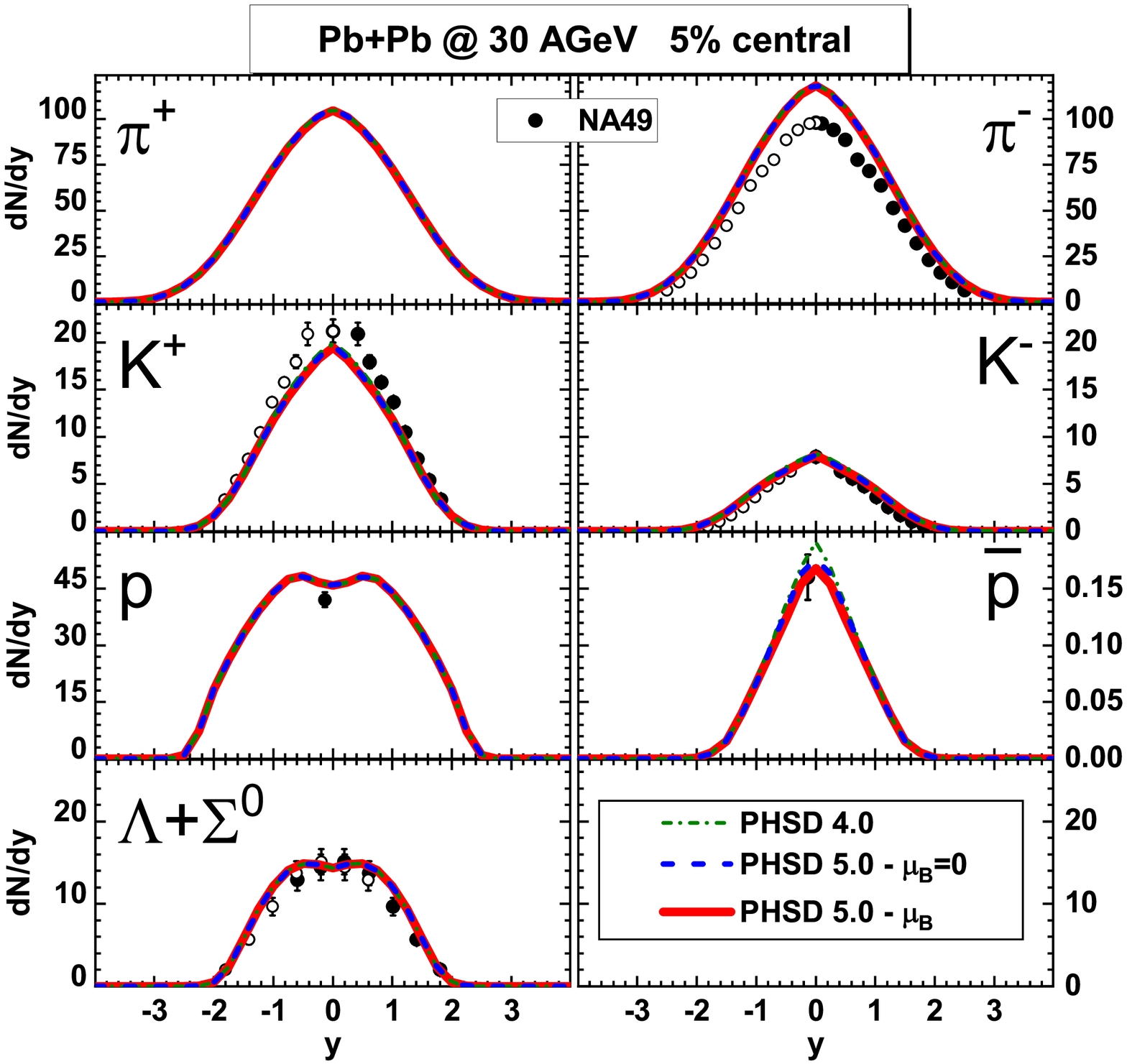}
	\caption{The rapidity distributions for  5\% central Pb+Pb collisions at 30 AGeV for PHSD4.0 (green dot-dashed lines), PHSD5.0 with partonic cross sections and parton masses calculated for $\mu_B$ = 0 (blue dashed lines) and with cross sections and parton masses evaluated at the actual chemical potential $\mu_B$ in each individual space-time cell (red lines) in comparison to the experimental data from the NA49 Collaboration \cite{115,116,117}. All PHSD results are practically the same within the linewidth.}
	\label{fig-dNdy-30AGeV}
\end{figure*}

\begin{figure*}[thb]
	\centering
	\includegraphics[width=110mm]{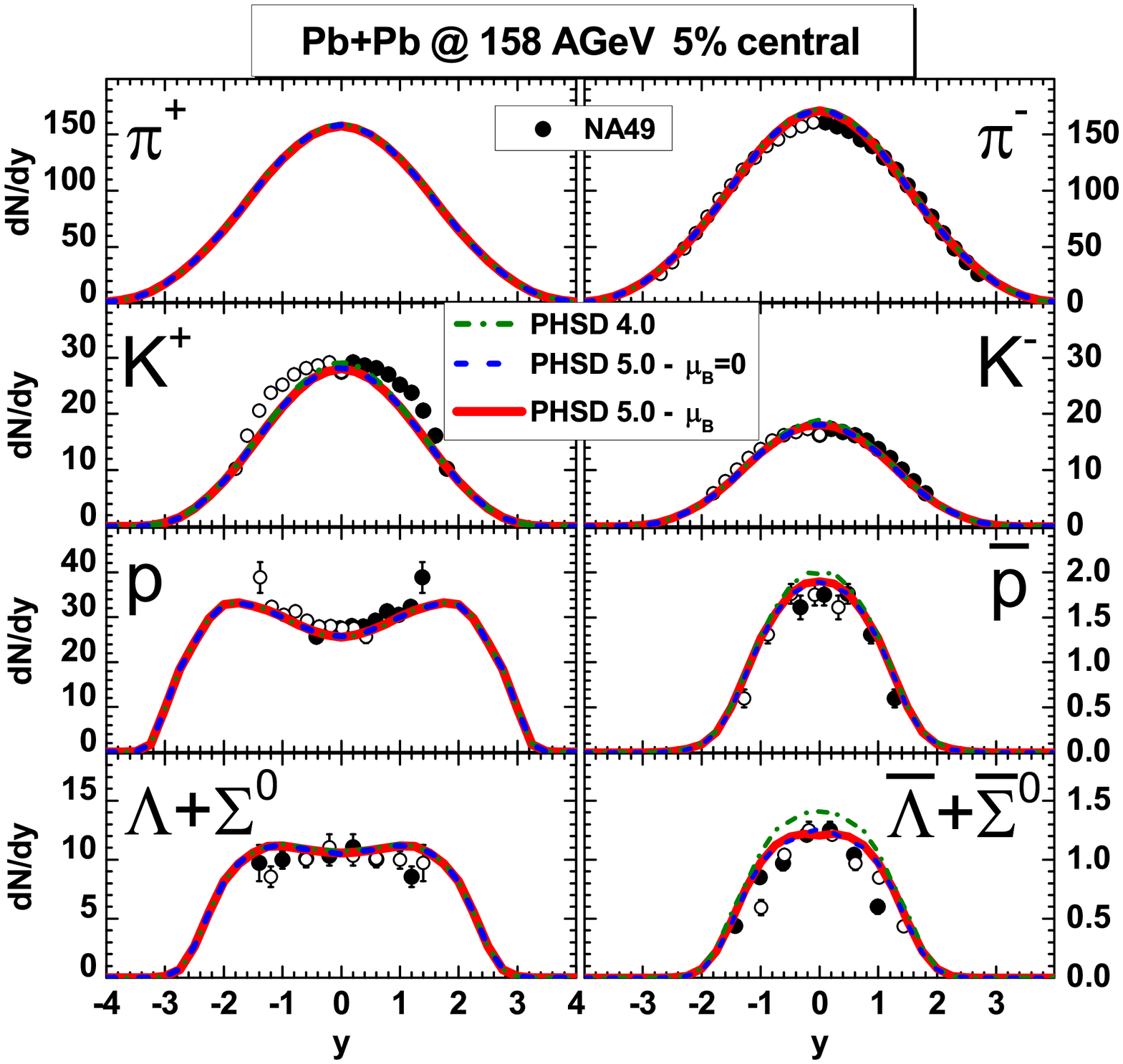}
	\caption{The rapidity distributions for  5\% central Pb+Pb collisions at 158 AGeV for PHSD4.0 (green dot-dashed lines), PHSD5.0 with partonic cross sections and parton masses calculated for $\mu_B$ = 0 (blue dashed lines) and with cross sections and parton masses evaluated at the actual chemical potential $\mu_B$ in each individual space-time cell (red lines) in comparison to the experimental data from the NA49 Collaboration \cite{118,119,120,121}. All PHSD results are the same within the linewidth except for the antibaryons. }
	\label{fig-dNdy-158AGeV}
\end{figure*}

Fig. \ref{fig-dNdy-11AGeV} displays the actual results for hadronic rapidity distributions in case of 5\% central Au+Au collisions at 10.7 AGeV for PHSD4.0 (green dot-dashed lines), PHSD5.0 with partonic cross sections and parton masses calculated for $\mu_B$ = 0 (blue dashed lines), and with cross sections and parton masses evaluated at the actual chemical potential $\mu_B$ in each individual space-time cell (red lines) in comparison to the experimental data from the E866 \cite{110}, E877 \cite{111}, E891 \cite{112}, E877 \cite{113} and E896 \cite{114} collaborations. Here we focus on the most abundant hadrons, i.e. pions, kaons, protons and neutral hyperons. We note in passing that the effects of chiral symmetry restoration are incorporated as in Ref. \cite{Palmese} since this was found to be mandatory to achieve a reasonable description of the strangeness degrees of freedom reflected in the kaon and neutral hyperon dynamics. As seen from Fig. \ref{fig-dNdy-11AGeV} there is no difference in rapidity distributions for all the hadron species from the different versions of PHSD within linewidth which implies that there is no sensitivity to the new partonic differential cross sections and parton masses employed. One could argue that this result might be due to the low amount of QGP produced at this energy but the different PHSD calculations for 5\% central Pb+Pb collisions at 30 AGeV in Fig. \ref{fig-dNdy-30AGeV} for the hadronic rapidity distributions do not provide a different picture, too. Only when stepping up to the top SPS energy of 158 AGeV one can identify a small difference in the antibaryon sector (${\bar p}$, ${\bar \Lambda} + {\bar \Sigma}^0$) in case of 5\% central Pb+Pb collisions (cf. Fig. \ref{fig-dNdy-158AGeV}).

\section{Summary}
\label{Section8}

In this contribution we have described the PHSD transport approach \cite{Review} and its recent extension to PHSD5.0 \cite{Pierre} to incorporate differential "off-shell" cross sections for all binary partonic channels that are based on the same effective propagators and couplings as employed in the QGP equation of state and the parton propagation. To this end we have recalled the extraction of the partonic masses and the coupling $g^2$ from lattice QCD data (within the DQPM) and calculated the partonic differential cross sections as a function of $T$ and $\mu_B$ for the leading tree-level diagrams (cf. Appendices of Ref. \cite{Pierre}). Furthermore,  we have used these differential cross sections to evaluate partonic scattering rates $\Gamma_i(T,\mu_B)$ for fixed $T$ and $\mu_B$ as well as to compute the ratio of the shear viscosity $\eta$ to entropy density $s$ within the Kubo formalism in comparison to calculations from lQCD.
It turns out that the ratio $\eta/s$ calculated with the partonic scattering rates in the relaxation-time approximation is very similar to the original result from the DQPM and to lQCD results such that the present extension of the PHSD approach does not lead to different partonic transport properties.
We recall that the novel PHSD version (PHSD5.0) is practically parameter free in the partonic sector since the effective coupling (squared) is determined by a fit to the scaled entropy density from lQCD. The dynamical masses for quarks and gluons then are fixed by the HTL expressions. The interaction rate in the time-like sector is, furthermore, calculated in leading order employing the DQPM propagators and coupling.

When implementing the differential cross sections and parton masses into the PHSD5.0 approach one has to specify the Lagrange parameters $T$ and $\mu_B$ in each computational cell in space-time. This has been done by employing the DQPM equation of state, which is practically identical to the lattice QCD equation of state,  and a diagonalization of the energy-momentum tensor from PHSD as described in Ref. \cite{Pierre}. 

In Section \ref{Section6} we then have calculated 5\% central Au+Au (or Pb+Pb) collisions and compared the results for hadronic rapidity distributions  from the previous PHSD4.0 with the novel version PHSD5.0 (with and without the explicit dependence of the partonic differential cross sections and parton masses on $\mu_B$).  No differences for all the hadron bulk observables from the various PHSD versions have been found at AGS and FAIR/NICA energies within linewidth which implies that there is no sensitivity to the new partonic differential cross sections employed. Only in case of the kaons and the antibaryons $\bar{p}$ and $\bar{\Lambda} + \bar{\Sigma}^0$, a small difference between PHSD4.0 and PHSD5.0 could be seen at top SPS energy, however, no clear difference between the PHSD5.0 calculations with partonic cross sections for $\mu_B$ = 0  and actual $\mu_B$ in the local cells.

Our findings can be understood as follows:
The fact that we find only small traces of the $\mu_B$-dependence of
partonic scattering dynamics in heavy-ion bulk observables - although the
differential cross sections and parton masses clearly depend on $\mu_B$ - means that one needs a sizable partonic density and large space-time QGP volume to explore the dynamics in the QGP phase. These conditions are only fulfilled at high bombarding energies (top SPS, RHIC energies) where, however, $\mu_B$ is rather low. On the other hand, decreasing the bombarding energy to FAIR/NICA energies and, thus, increasing $\mu_B$, leads to collisions that are dominated by the hadronic phase where the extraction of information about the parton dynamics will be rather complicated based on bulk observables. Further investigations of other observables (such as flow coefficients $v_n$ of particles and antiparticles, fluctuations and correlations) might contain more visible $\mu_B-$traces from the QGP phase.

\begin{acknowledgement}
The authors acknowledge inspiring discussions with J. Aichelin, H. Berrehrah, C. Ratti, E. Seifert, A. Palmese and T. Steinert. This work was supported  by the LOEWE center "HIC for FAIR". Furthermore, P.M., L.O. and E.B. acknowledge support by the Deutsche Forschungsgemeinschaft (DFG, German Research Foundation) through the grant CRC-TR 211 'Strong-interaction matter under extreme conditions' - Project number 315477589 - TRR 211. O.S. acknowledges support from HGS-HIRe for FAIR; L.O. and E.B. thank the COST Action THOR, CA15213.
The computational resources have been provided by the LOEWE-Center for Scientific Computing.
\end{acknowledgement}


\begin{thebibliography}{99}
%
\bibitem{Cugnon}
J. Cugnon, Phys. Rev. C22, 1885 (1980).

\bibitem{UU} 
E. A. Uehling  and  G. E. Uhlenbeck, Phys. Rev. 43, 552 (1932).
\bibitem{Bertsch} 	
G. F. Bertsch and S. Das Gupta, Phys. Rept. 160, 189 (1988).

\bibitem{Cas90}
W. Cassing, V. Metag, U. Mosel, and K. Niita,
Phys. Rept. 188, 363  (1990).

\bibitem{Aichelin}
J.Aichelin  and H. St\"ocker, Phys. Lett. B176, 14 (1986).
\bibitem{Hartnack} 
C. Hartnack, K. Puri Rajeev, J. Aichelin , Konopka J., S.A. Bass, Horst St\"ocker, and W. Greiner, Eur. Phys. J. A1, 151-169 (1998).

\bibitem{Ko}
C.M. Ko, Q. Li, and Ren-Chuan Wang,
Phys.  Rev. Lett. 59, 1084 (1987).

\bibitem{Blaettel}
B. Blaettel, V. Koch, W. Cassing, and U. Mosel,
 Phys. Rev. C38, 1767 (1988).
 \bibitem{Maruyama}
Tomoyuki Maruyama, Wolfgang Cassing, Ulrich Mosel, Stefan Teis, Klaus Weber,
 Nucl.Phys. A573, 653-675 (1994).

\bibitem{JETSET}
      H.-U. Bengtsson and T. Sj\"ostrand,
      Comp. Phys. Commun. {46}, 43 (1987).

\bibitem{Ehehalt} W. Ehehalt and W. Cassing. Nucl. Phys. A602, 449 (1996).

\bibitem{HSD}
      W. Cassing and E. L. Bratkovskaya, Phys. Rep. 308, 65 (1999).

\bibitem{UrQMD} S. A. Bass et al.,  Prog. Part. Nucl. Phys. 41, 255-369 (1998).


\bibitem{Brat2004}
E.L. Bratkovskaya, M. Bleicher, M. Reiter, S. Soff, H. St\"ocker, M. van Leeuwen, S.A. Bass, W. Cassing,
 Phys. Rev. C69, 054907 (2004).
\bibitem{Marty} 
R. Marty and J. Aichelin, Phys. Rev. C87, 034912 (2013).

\bibitem{Marty2} 
R. Marty, E. Bratkovskaya, W. Cassing, and  J. Aichelin,  Phys. Rev. C92, 015201 (2015).



\bibitem{Sascha1}
      S. Juchem, W. Cassing and C. Greiner, Phys. Rev. D 69, 025006 (2004);
      Nucl. Phys. A 743, 92 (2004).

\bibitem{Juchem}
      W. Cassing and S. Juchem, Nucl. Phys. A 665 (2000) 377;
      {\it ibid} A 672, 417 (2000).

\bibitem{Knoll1}
      Y. B. Ivanov, J. Knoll and D. N. Voskresensky,
      Nucl. Phys. A 672, 313 (2000).
      
\bibitem{Brat2008}
E.L. Bratkovskaya and  W. Cassing,  Nucl. Phys. A807, 214 (2008).


\bibitem{Shuryak}
      E. Shuryak, Prog. Part. Nucl. Phys. 53, 273 (2004).

\bibitem{Thoma} M. H. Thoma, J. Phys. G 31, L7 (2005); Nucl. Phys. A 774, 307 (2006).

\bibitem{Andre}
      A. Peshier and W. Cassing, Phys. Rev. Lett. 94, 172301 (2005).


\bibitem{STARS}
       I. Arsene {\it et al.}, Nucl. Phys. A 757, 1 (2005);
       B. B. Back {\it et al.}, Nucl. Phys. A 757, 28 (2005);
       J. Adams {\it et al.}, Nucl. Phys. A 757, 102 (2005);
       K. Adcox {\it et al.}, Nucl. Phys. A 757 184, (2005).

\bibitem{Miklos3}
      T. Hirano and M. Gyulassy, Nucl. Phys. A 769, 71 (2006).


\bibitem{Koal1}
      R. C. Hwa and C. B. Yang, Phys. Rev. C 67, 034902 (2003);
      V. Greco, C. M. Ko and P. Levai, Phys. Rev. Lett. 90, 202302 (2003).

\bibitem{Koal2}
      R. J. Fries, B. M\"uller, C. Nonaka and S. A. Bass,
      Phys. Rev. Lett. 90, 202303 (2003).

\bibitem{AMPT} Z.-W. Lin {\it et al.}, Phys. Rev. C 72, 064901 (2005).
\bibitem{PRC08} W. Cassing and E. L. Bratkovskaya, Phys. Rev. C 78, 034919 (2008).


\bibitem{Cassing06}
      W. Cassing, Nucl. Phys. A 791, 365 (2007).

\bibitem{Cassing07}
      W. Cassing, Nucl. Phys. A 795, 70 (2007).
      
\bibitem{Crev} W. Cassing, E. Phys. J. ST 168, 3 (2009).

\bibitem{Review}
O. Linnyk, E.L. Bratkovskaya, and W. Cassing, Prog. Part. Nucl. Phys. 87, 50 (2016).


\bibitem{CaBra09} W. Cassing and E. L. Bratkovskaya, { Nucl. Phys.} A {831}, 215 (2009).
\bibitem{Moreau} 
P. Moreau, O. Linnyk, W. Cassing, and E. Bratkovskaya, Phys. Rev. C93, 044916 (2016).
\bibitem{Ozvenchuk13}
 V. Ozvenchuk, O. Linnyk, M. I. Gorenstein, E. L. Bratkovskaya, and W. Cassing, Phys. Rev. C 87, 024901 (2013).


\bibitem{Pierre}
P. Moreau, O. Soloveva, L. Oliva, T. Song, W. Cassing, and E. Bratkovskaya, Phys. Rev. C 100, 014911 (2019).

\bibitem{85} 
R. Kubo, J. Phys. Soc. Jpn. 12, 570 (1957).
\bibitem{93} 
V. Ozvenchuk, O. Linnyk, M. I. Gorenstein, E. L. Bratkovskaya, and W. Cassing, Phys. Rev. C 87, 064903 (2013).
\bibitem{87} G. Aarts and J. M. Martinez Resco, J. High Energy Phys. 04, 053 (2002).
\bibitem{89} R. Lang, N. Kaiser, and W. Weise, Eur. Phys. J. A 48, 109 (2012).
\bibitem{94} C. Sasaki and K. Redlich, Phys. Rev. C 79, 055207 (2009).
\bibitem{21} J.-P. Blaizot and E. Iancu, Nucl. Phys. B 557, 183 (1999).
\bibitem{100} P. K. Kovtun, D. T. Son, and A. O. Starinets, Phys. Rev. Lett. 94, 111601 (2005).
\bibitem{101} N. Astrakhantsev, V. Braguta, and A. Kotov, J. High Energy
Phys. 04, 101 (2017).
\bibitem{Palmese}  A. Palmese, W. Cassing, E. Seifert, T. Steinert, P. Moreau, and
E. L. Bratkovskaya, Phys. Rev. C 94, 044912 (2016).

\bibitem{110} Y. Akiba et al. (E802 Collaboration), Nucl. Phys. A 610, 139
(1996). 
\bibitem{111} R. Lacasse et al. (E877 Collaboration), Nucl. Phys. A 610, 153
(1996).
\bibitem{112} S. Ahmad et al., Phys. Lett. B 382, 35 (1996); 386, 496(E)
(1996).
\bibitem{113} J. Barrette et al. (E877 Collaboration), Phys. Rev. C 63,
014902 (2001).
\bibitem{114} S. Albergo et al., Phys. Rev. Lett. 88, 062301 (2002).
\bibitem{115} C. Alt et al. (NA49 Collaboraion), Phys. Rev. C 73, 044910
(2006).
\bibitem{116} C. Alt et al. (NA49 Collaboration), Phys. Rev. C 77, 024903
(2008).
\bibitem{117} C. Alt et al. (NA49 Collaboration), Phys. Rev. C 78, 034918
(2008).
\bibitem{118} S. V. Afanasiev et al. (NA49 Collaboration), Phys. Rev. C 66,
054902 (2002).
\bibitem{119} T. Anticic et al. (NA49 Collaboration), Phys. Rev. Lett. 93,
022302 (2004).
\bibitem{120} T. Anticic et al. (NA49 Collaboration), Phys. Rev. C 83,
014901 (2011).
\bibitem{121} T. Anticic et al. (NA49 Collaboration), Phys. Rev. C 86,
054903 (2012).

\end{thebibliography}
\end{document}